\documentclass[twocolumn,letter]{jpsj3}
\usepackage{graphicx}
\usepackage{color}
\usepackage{bm}

\title{%
Bulk--Boundary Correspondence in a Non-Hermitian Chern Insulator
}

\author{%
Yositake Takane
}

\inst{%
Department of Quantum Matter,
Graduate School of Advanced Science and Engineering,\\
Hiroshima University, Higashihiroshima, Hiroshima 739-8530, Japan
}

\recdate{ \hspace{50mm} }

\abst{%
A scenario of non-Hermitian bulk--boundary correspondence proposed
for one-dimensional topological insulators
is adapted to a non-Hermitian Chern insulator
to examine its applicability to two-dimensional systems.
This scenario employs
bulk geometry under a modified periodic boundary condition
and boundary geometry under an open boundary condition.
The bulk geometry is used to define a topological number,
whereas the boundary geometry is used to observe
the presence or absence of a topological boundary state.
It is demonstrated that the bulk--boundary correspondence holds
in a two-dimensional Chern insulator with gain/loss-type non-Hermiticity;
a nontrivial Chern number calculated in the bulk geometry is
in one-to-one correspondence with
the presence of a topological boundary state in the boundary geometry.
This approach enables us to determine a phase diagram in the boundary geometry.
}

\begin{document}
\maketitle

Bulk--boundary correspondence is a characteristic feature revealed in
topological systems.~\cite{thouless,kohmoto,kane,fu1,moore,roy,ryu1,fu2,
hasan,qi}
Its original scenario proposed by Hatsugai and co-worker~\cite{hatsugai,ryu2}
employs bulk geometry under a periodic boundary condition (pbc)
and boundary geometry under an open boundary condition (obc).
The former is used to define a topological number
and the latter is used to observe the presence or absence of
a topological boundary state.
A nontrivial topological number is in one-to-one
correspondence with the presence of a topological boundary state.

Studies on topological systems are now extended to
the non-Hermitian regime,~\cite{hatano,bender,brody} covering a variety of
non-Hermitian topological systems.~\cite{esaki,t_lee,xu,ashida,xiong,gong,
kunst1,yao1,yao2,zyuzin,alvarez,yoshida1,kawabata1,kawabata2,longhi1,
yokomizo1,okuma1,song,herviou,okugawa,yoshida2,c_lee,papaj,kunst2,imura1,zhou1,
torres,borgnia,okuma2,zhang,yi,yang,yokomizo2,zhou2,e_lee,wu1,wu2,longhi2,
kawabata3,mochizuki,li,koch,yuce,mostafavi,wang,zhao,he,yokomizo3,
imura2,kondo,kawasaki,yoshida3}
The idea of bulk--boundary correspondence is also extended to
non-Hermitian topological systems.
Previous studies showed that this correspondence
is broken in some cases.~\cite{t_lee,xiong}
This should be closely related to a non-Hermitian skin effect;~\cite{yao1}
eigenfunctions in a non-Hermitian system under the obc are localized
near a boundary of the system owing to non-Hermiticity.
A possible explanation of the breaking of bulk--boundary correspondence
is as follows.
In a non-Hermitian topological system under the obc, eigenfunctions tend to
localize near a boundary as a result of the non-Hermitian skin effect.
In contrast, all eigenfunctions under the pbc extend over the entire system.
Therefore, a topological number defined in the bulk geometry under the pbc
never reflects the non-Hermitian skin effect, and hence
it fails to describe the topological nature of the system under the obc.

To restore the bulk--boundary correspondence
in non-Hermitian topological systems,
several scenarios have been proposed.~\cite{yao1,yokomizo1,kunst1,borgnia}
However, none of these share the original idea
in the sense that they consider only the boundary geometry.
Recently, another scenario of bulk--boundary correspondence has been proposed
in Refs.~\citen{imura1} and \citen{imura2}.
This scenario employs bulk and boundary geometries as in the original one.
A crucial point is that a modified periodic boundary condition
(mpbc) is imposed on the bulk geometry instead of the pbc.
The mpbc for a wavefunction $\psi(n)$ in a one-dimensional lattice system
of length $L$ is given by $\psi(L+n) = b^{L}\psi(n)$,
where $b$ is a positive real constant.~\cite{imura1}
A wavefunction $\psi(n) = \beta^{n}$ satisfies this boundary condition
when $\beta = b e^{ika}$, where $a$ is the lattice constant and
$k = 2\pi l/L$ with $l = 0,1,2,\dots,L-1$.
If $b > 1$ ($b < 1$), the mpbc selects wavefunctions
that exponentially increase (decrease) with increasing $n$.
This behavior of $\psi(n)$ is essentially equivalent to
a non-Hermitian skin effect.
Therefore, by using the bulk geometry under the mpbc, we can take
the non-Hermitian skin effect into account
in calculating a topological number.
The scenario in Refs.~\citen{imura1} and \citen{imura2}
succeeds in describing the bulk--boundary correspondence
in one-dimensional non-Hermitian topological insulators
by treating $b$ as an additional parameter of the system.
An advantage of this approach is that, at the cost of introducing $b$,
we can demonstrate the bulk--boundary correspondence in non-Hermitian systems
in almost the same way as in Hermitian systems.
However, its applicability has not been examined
in two- and three-dimensional systems.

In this Letter, we adapt the scenario in Refs.~\citen{imura1}
and \citen{imura2} to a Chern insulator with gain/loss-type non-Hermiticity
to examine its applicability to two-dimensional systems.
It is demonstrated that this scenario properly describes
the bulk--boundary correspondence in the non-Hermitian Chern insulator
and enables us to determine a phase diagram in the boundary geometry.

We adopt a tight-binding model for non-Hermitian Chern insulators
on a square lattice with lattice constant $a$.
The Hamiltonian is given by
$H = H_{\rm d}+H_{x}+H_{y}$ with~\cite{yao2}
\begin{align}
   H_{\rm d}
 & = \sum_{m,n} |m,n \rangle
                  \left[ 
                    \begin{array}{cc}
                      M & (i+1)\gamma \\
                      (i-1)\gamma & -M \\
                    \end{array}
                  \right]
                \langle m,n| ,
         \\
   H_{x}
 & = \sum_{m,n} |m+1,n \rangle
                  \left[ 
                    \begin{array}{cccc}
                      -\frac{1}{2}t & \frac{1}{2}iv \\
                      \frac{1}{2}iv & \frac{1}{2}t \\
                    \end{array}
                  \right]
                \langle m,n|
              + {\rm h.c.} ,
        \\
   H_{y}
 & = \sum_{m,n} |m,n+1 \rangle
                  \left[ 
                    \begin{array}{cccc}
                      -\frac{1}{2}t & \frac{1}{2}v \\
                      -\frac{1}{2}v & \frac{1}{2}t \\
                    \end{array}
                  \right]
                \langle m,n|
              + {\rm h.c.} ,
\end{align}
where $m$ and $n$ respectively specifies the location of each site
in the $x$- and $y$-directions
and $|m,n \rangle$ consists of spin up and down components:
$|m,n \rangle
= \bigl\{ |m,n \rangle_{\uparrow}, |m,n \rangle_{\downarrow} \bigr\}$.
The parameters are set in a symmetric manner
with respect to the $x$- and $y$-directions.
The non-Hermiticity is introduced by the terms with $\gamma$.
For simplicity, we focus on the case of
$M \ge 0$ and $\gamma \ge 0$ with $v = t > 0$.
As relevant parameters, we hereafter use
\begin{align}
        \label{eq:def_rel-para}
  \tilde{M} = \frac{M}{t}, \hspace{10mm} \tilde{\gamma} = \frac{\gamma}{t} .
\end{align}
This model becomes topologically nontrivial when $\tilde{M} < 2$
in the Hermitian limit of $\tilde{\gamma} = 0$.

Let us first consider the model
in the bulk geometry of $L \times L$ sites under an mpbc.
In accordance with the symmetric setting of the parameters,
we are allowed to impose the following mpbc on a wavefunction
$\psi_{\sigma}(m,n)$:
\begin{align}
      \label{eq:BC_mpbc}
 \psi_{\sigma}(m+L,n) = \psi_{\sigma}(m,n+L) = b^{L}\psi_{\sigma}(m,n) ,
\end{align}
where $\sigma = \uparrow$, $\downarrow$.
A plane-wave-like solution compatible with Eq.~(\ref{eq:BC_mpbc})
is given by $\psi_{\sigma}(m,n) = \beta_{x}^{m}\beta_{y}^{n}$ with
\begin{align}
  \beta_{x} = be^{ik_{x}a}, \hspace{3mm} \beta_{y} = be^{ik_{y}a} ,
\end{align}
where $k_{i} = 2n_{i}\pi/L$ and $n_{i} = 0,1,2,\dots, L-1$ ($i = x, y$).
With these basis functions, we can transform $H$ as
\begin{align}
      \label{eq:H(k)-mpbc}
   H(k_{x},k_{y},b)
 & = t
     \left[ 
       \begin{array}{cc}
         \eta_{z} & \eta_{x}-i\eta_{y} \\
         \eta_{x}+i\eta_{y} & -\eta_{z} \\
       \end{array}
     \right] ,
\end{align}
with
\begin{align}
 & \eta_{x}(k_{x},b)
   = b_{+}\sin(k_{x}a) +i\left[-b_{-}\cos(k_{x}a) +\tilde{\gamma} \right] ,
      \\
 & \eta_{y}(k_{y},b)
  = b_{+}\sin(k_{y}a) +i\left[-b_{-}\cos(k_{y}a) +\tilde{\gamma} \right] ,
      \\
 & \eta_{z}(k_{x},k_{y},b)
   = \tilde{M} - b_{+}\left[\cos(k_{x}a)+\cos(k_{y}a)\right]
      \nonumber \\
 & \hspace{20mm}
     -ib_{-}\left[ \sin(k_{x}a) + \sin(k_{x}a) \right] ,
\end{align}
where $v = t$ is assumed and
\begin{align}
    b_{\pm} = \frac{1}{2}\left(b \pm b^{-1}\right) .
\end{align}
The energy (normalized by $t$) of an eigenstate characterized by
$\mib{k}=(k_{x},k_{y})$ and $b$ is given by $E = \pm \epsilon$ with
$\epsilon = \sqrt{\eta_{x}^{2}+\eta_{y}^{2}+\eta_{z}^{2}}$,
where ${\rm Re}\{\epsilon\} \ge 0$.

Let $\langle\psi_{\mib{k}}^{L}|$ and $|\psi_{\mib{k}}^{R}\rangle$
respectively be the left and right eigenstates of $H$
with the eigenvalue of $-\epsilon$,
satisfying $\langle\psi_{\mib{k}}^{L}|\psi_{\mib{k}}^{R}\rangle = 1$.
They are expressed as
\begin{align}
  \langle\psi_{\mib{k}}^{L}| = \frac{1}{\sqrt{A}}
                    ^{t}\!
                    \left[ \begin{array}{c}
                              \eta_{z}-\epsilon \\
                              \eta_{x}-i\eta_{y}
                     \end{array}
                    \right] ,
  \hspace{2mm}
  |\psi_{\mib{k}}^{R}\rangle = \frac{1}{\sqrt{A}}
                    \left[ \begin{array}{c}
                              \eta_{z}-\epsilon \\
                              \eta_{x}+i\eta_{y}
                     \end{array}
                    \right] ,
\end{align}
where $A = \eta_{x}^{2}+\eta_{y}^{2}+(\eta_{z}-\epsilon)^{2}$.
When the two bands of $E = \pm \epsilon$ are separated
by a gap (i.e., ${\rm Re}\{\epsilon\} \neq 0$ for an arbitrary $\mib{k}$),
the Chern number $N$ is defined as~\cite{yao2}
\begin{align}
          \label{eq:def-N}
  N = \frac{1}{2\pi i}\int d^{2}k
      \left( \frac{\partial \langle\psi_{\mib{k}}^{L}|}{\partial k_{x}}
             \frac{\partial |\psi_{\mib{k}}^{R}\rangle}{\partial k_{y}}
           - \frac{\partial \langle\psi_{\mib{k}}^{L}|}{\partial k_{y}}
             \frac{\partial |\psi_{\mib{k}}^{R}\rangle}{\partial k_{x}}
      \right) ,
\end{align}
which is expressed as
\begin{align}
  N = \int \frac{d^{2}k}{4 \pi} \frac{1}{\epsilon^{3}}
      \left(  \eta_{x}\frac{\partial \eta_{y}}{\partial k_{y}}
              \frac{\partial \eta_{z}}{\partial k_{x}}
            + \frac{\partial \eta_{x}}{\partial k_{x}}
              \eta_{y}\frac{\partial \eta_{z}}{\partial k_{y}}
            - \frac{\partial \eta_{x}}{\partial k_{x}}
              \frac{\partial \eta_{y}}{\partial k_{y}}\eta_{z}
      \right) .
\end{align}
Here, the limit of $L \to \infty$ is implicitly assumed.
The Chern number $N$ takes an integer, which depends on
not only $\tilde{M}$ and $\tilde{\gamma}$ but also $b$.
In the Hermitian limit of $\tilde{\gamma} = 0$,
Eq.~(\ref{eq:def-N}) at $b = 1$ becomes equivalent to
an ordinary expression of the Chern number.~\cite{thouless,kohmoto}

The system under consideration takes three phases:
a topologically trivial phase of $N = 0$, a nontrivial phase of $N = 1$,
and a gapless phase, in which $N$ cannot be defined.
Below, we consider $N$ for a given $\tilde{M}$
in a parameter space of $\tilde{\gamma}$ and $b$, where
these phases are separated by lines on which the gap closes.
Let us find such gap closing lines.
In this system, a gap closing takes place when
${\rm Re}\{\epsilon\} = {\rm Im}\{\epsilon\} = 0$.
The expression of $\epsilon^{2} = \eta_{x}^{2}+\eta_{y}^{2}+\eta_{z}^{2}$
indicates that ${\rm Im}\{\epsilon^{2}\} = 0$ for
$\mib{k} =(0,0)$, $(0,\frac{\pi}{a})$, $(\frac{\pi}{a},0)$,
and $(\frac{\pi}{a},\frac{\pi}{a})$,
where $\mib{k} = (\frac{\pi}{a},\frac{\pi}{a})$ can be ignored
because this point has a gap in relevant situations.
For $\mib{k}=(0,0)$, we find that ${\rm Re}\{\epsilon^{2}\} = 0$ when
$(\tilde{M}-2b_{+})^{2} - 2(-b_{-}+\tilde{\gamma})^{2} = 0$.
This equation yields four solutions of $b$:
\begin{align}
      \label{eq:b_1}
  b_{\rm 1} = \frac{\tilde{M}-\sqrt{2}\tilde{\gamma}
                    + \sqrt{(\tilde{M}-\sqrt{2}\tilde{\gamma})^{2}-2}}
                   {2-\sqrt{2}} ,
         \\
      \label{eq:b_2}
  b_{\rm 2} = \frac{\tilde{M}+\sqrt{2}\tilde{\gamma}
                    + \sqrt{(\tilde{M}+\sqrt{2}\tilde{\gamma})^{2}-2}}
                   {2+\sqrt{2}} ,
         \\
      \label{eq:b_3}
  b_{\rm 3} = \frac{\tilde{M}-\sqrt{2}\tilde{\gamma}
                    - \sqrt{(\tilde{M}-\sqrt{2}\tilde{\gamma})^{2}-2}}
                   {2-\sqrt{2}} ,
         \\
      \label{eq:b_4}
  b_{\rm 4} = \frac{\tilde{M}+\sqrt{2}\tilde{\gamma}
                    - \sqrt{(\tilde{M}+\sqrt{2}\tilde{\gamma})^{2}-2}}
                   {2+\sqrt{2}} .
\end{align}
If these solutions are positive real
and satisfy $b_{4} < b_{3} < b_{2} < b_{1}$,
the system has a gap at $\mib{k}=(0,0)$
if $b < b_{4}$, $b_{3} < b < b_{2}$, or $b_{1} < b$.
For $\mib{k} =(0,\frac{\pi}{a})$ and $(\frac{\pi}{a},0)$,
we find that ${\rm Re}\{\epsilon^{2}\} = 0$ when
$2b_{-}^{2} -\tilde{M}^{2} + 2\tilde{\gamma}^{2} = 0$.
This equation yields two solutions of $b$:
\begin{align}
      \label{eq:b_5}
  b_{5} & =  \sqrt{\frac{\tilde{M}^{2}}{2}-\tilde{\gamma}^{2}+1}
             + \sqrt{\frac{\tilde{M}^{2}}{2}-\tilde{\gamma}^{2}} ,
         \\
      \label{eq:b_6}
  b_{6} & = \sqrt{\frac{\tilde{M}^{2}}{2}-\tilde{\gamma}^{2}+1}
            - \sqrt{\frac{\tilde{M}^{2}}{2}-\tilde{\gamma}^{2}} .
\end{align}
If these solutions are real, the system has a gap
at $\mib{k}=(0,\frac{\pi}{a})$ and $(\frac{\pi}{a},0)$ if $b_{6} < b < b_{5}$.
Each $b_{i}$ ($i = 1, \dots, 6$) is irrelevant if it becomes a complex number.
In addition to $b_{i}$ ($i = 1, \dots, 6$),
we need to consider $b_{0}$ given below.
When $(\tilde{M}-\sqrt{2}\tilde{\gamma})^{2} < 2$, $b_{1}$ and $b_{3}$
become the complex numbers $b_{1} = b_{0}e^{ik_{0}a}$ and
$b_{3} = b_{0}e^{-ik_{0}a}$,
where $b_{0} = 1+\sqrt{2}$ and $k_{0}$ is defined by
\begin{align}
  \tan\left(k_{0}a\right)
   = \frac{\sqrt{2-(\tilde{M}-\sqrt{2}\tilde{\gamma})^{2}}}
                    {\tilde{M}-\sqrt{2}\tilde{\gamma}} .
\end{align}
We can show that ${\rm Re}\{\epsilon^{2}\} = {\rm Im}\{\epsilon^{2}\} = 0$
at $b = b_{0}$ if $\mib{k}=\pm (k_{0},k_{0})$.
Therefore, $b_{0}$ represents a gap closing line.

The phase diagrams in the bulk geometry for $\tilde{M} = 0.5$, $1.2$,
$2.2$, $2.5$, and $3.5$ are shown in Figs.~1(a)--1(e)
in the $\tilde{\gamma} b$-plane.
In each panel of Fig.~1, topologically trivial and nontrivial regions
are respectively designated as $N = 0$ and $N = 1$,
and the outside of these regions is a gapless region,
in which $N$ cannot be defined.
\begin{figure}[btp]
\begin{tabular}{cc}
\begin{minipage}{0.5\hsize}
\begin{center}
\hspace{-10mm}
\includegraphics[height=4.0cm]{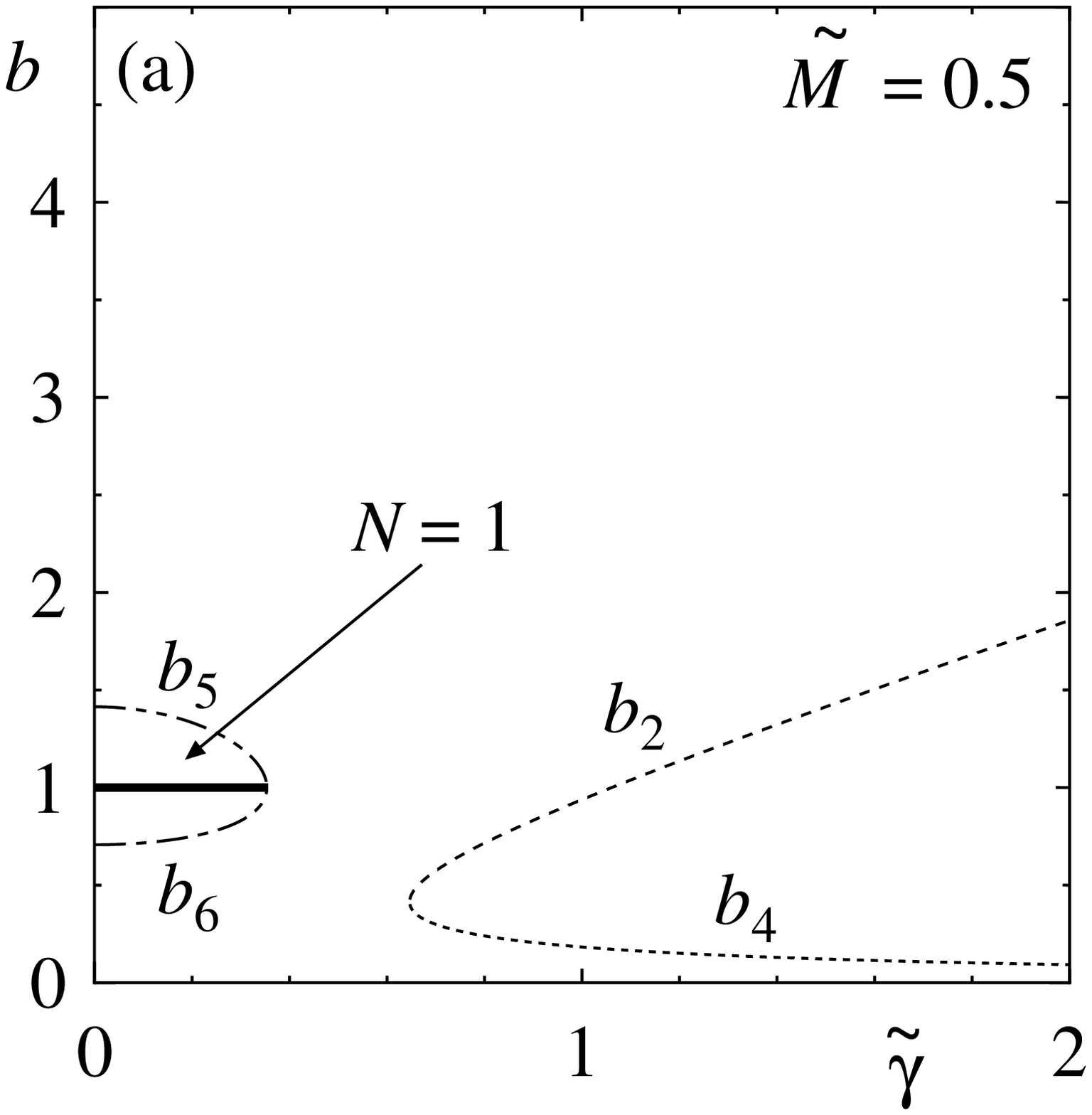}
\end{center}
\end{minipage}
\begin{minipage}{0.5\hsize}
\begin{center}
\hspace{-5mm}
\includegraphics[height=4.0cm]{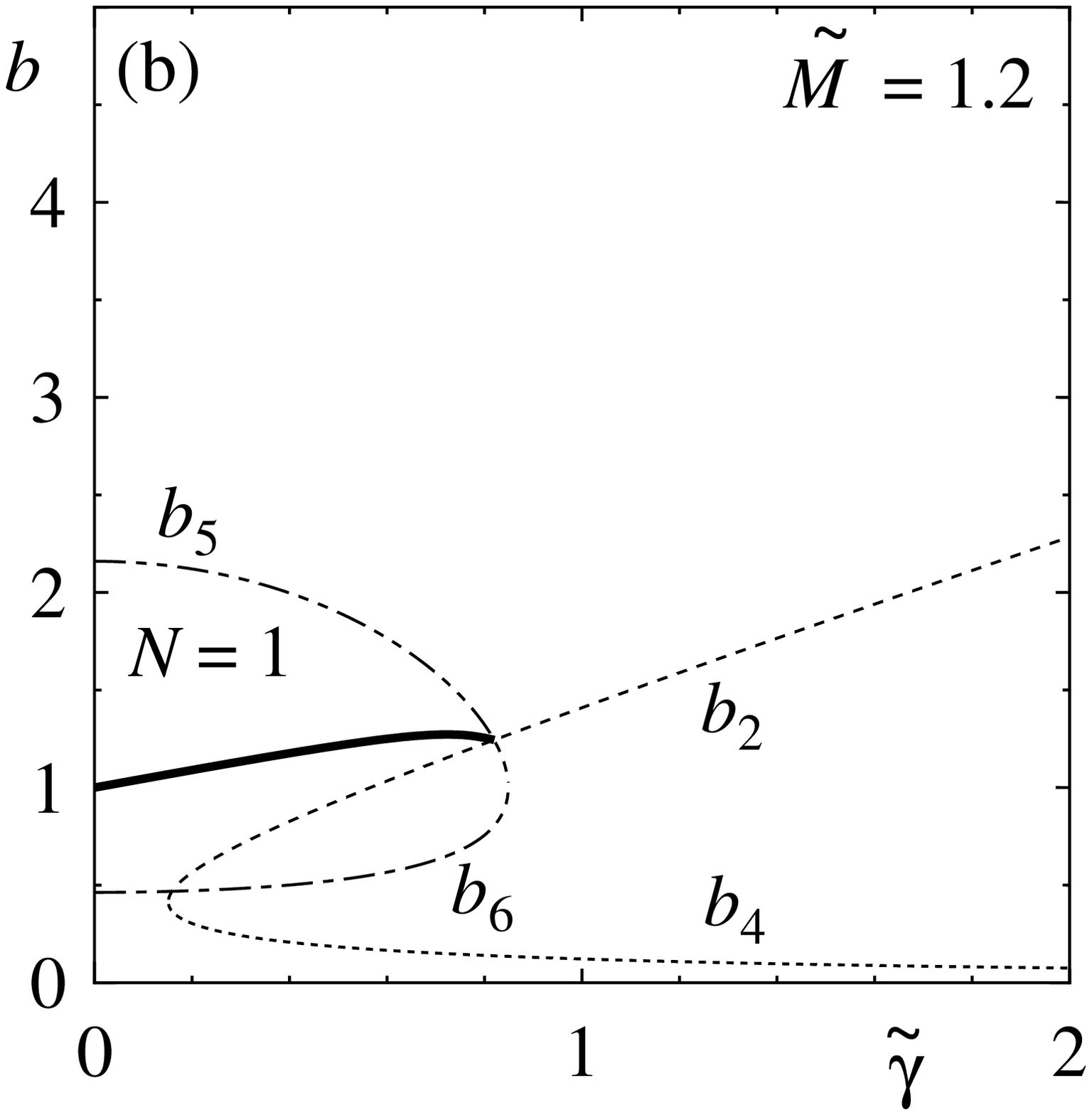}
\end{center}
\end{minipage}
\end{tabular}
\begin{tabular}{cc}
\begin{minipage}{0.5\hsize}
\begin{center}
\hspace{-10mm}
\includegraphics[height=4.0cm]{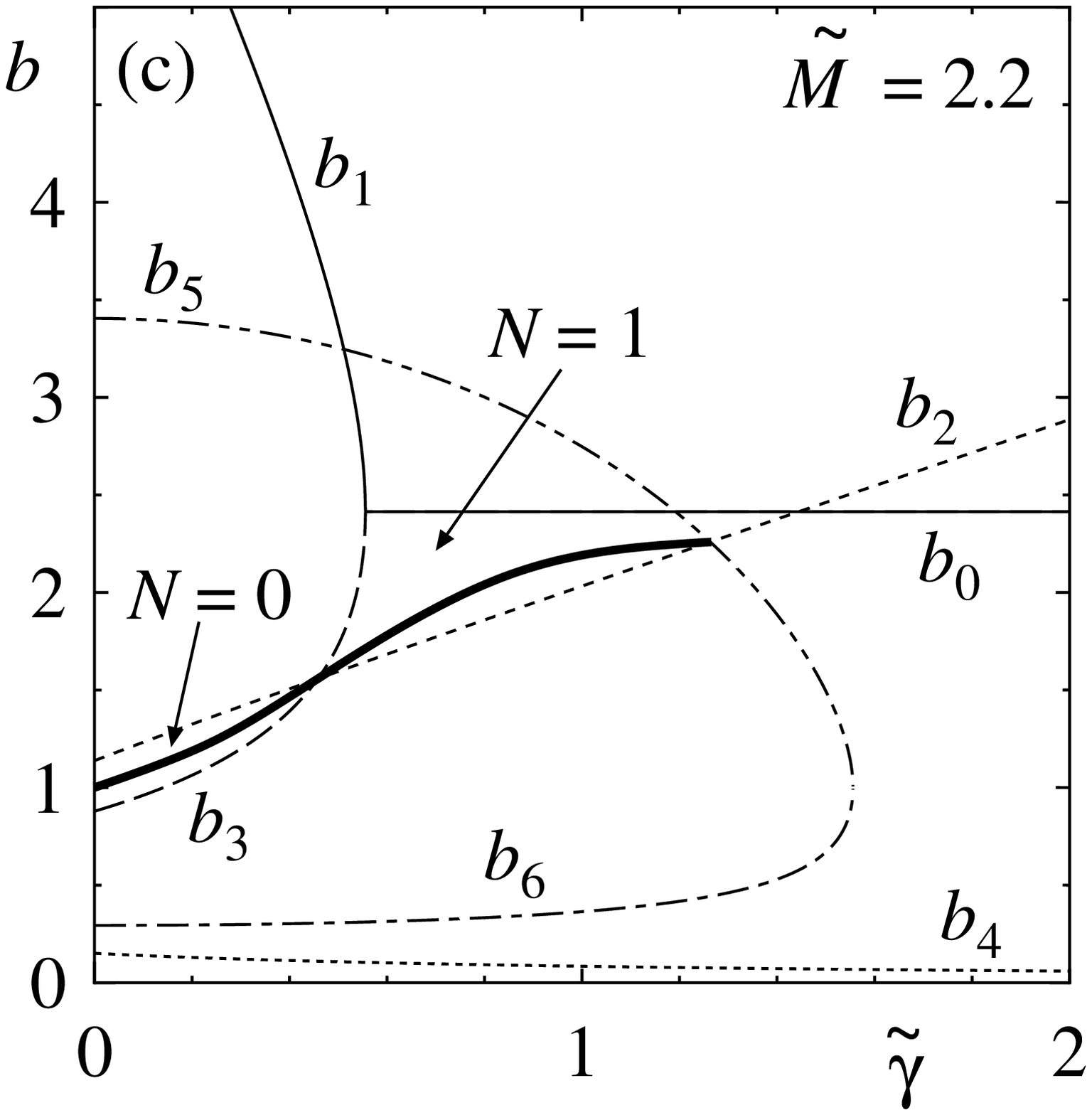}
\end{center}
\end{minipage}
\begin{minipage}{0.5\hsize}
\begin{center}
\hspace{-5mm}
\includegraphics[height=4.0cm]{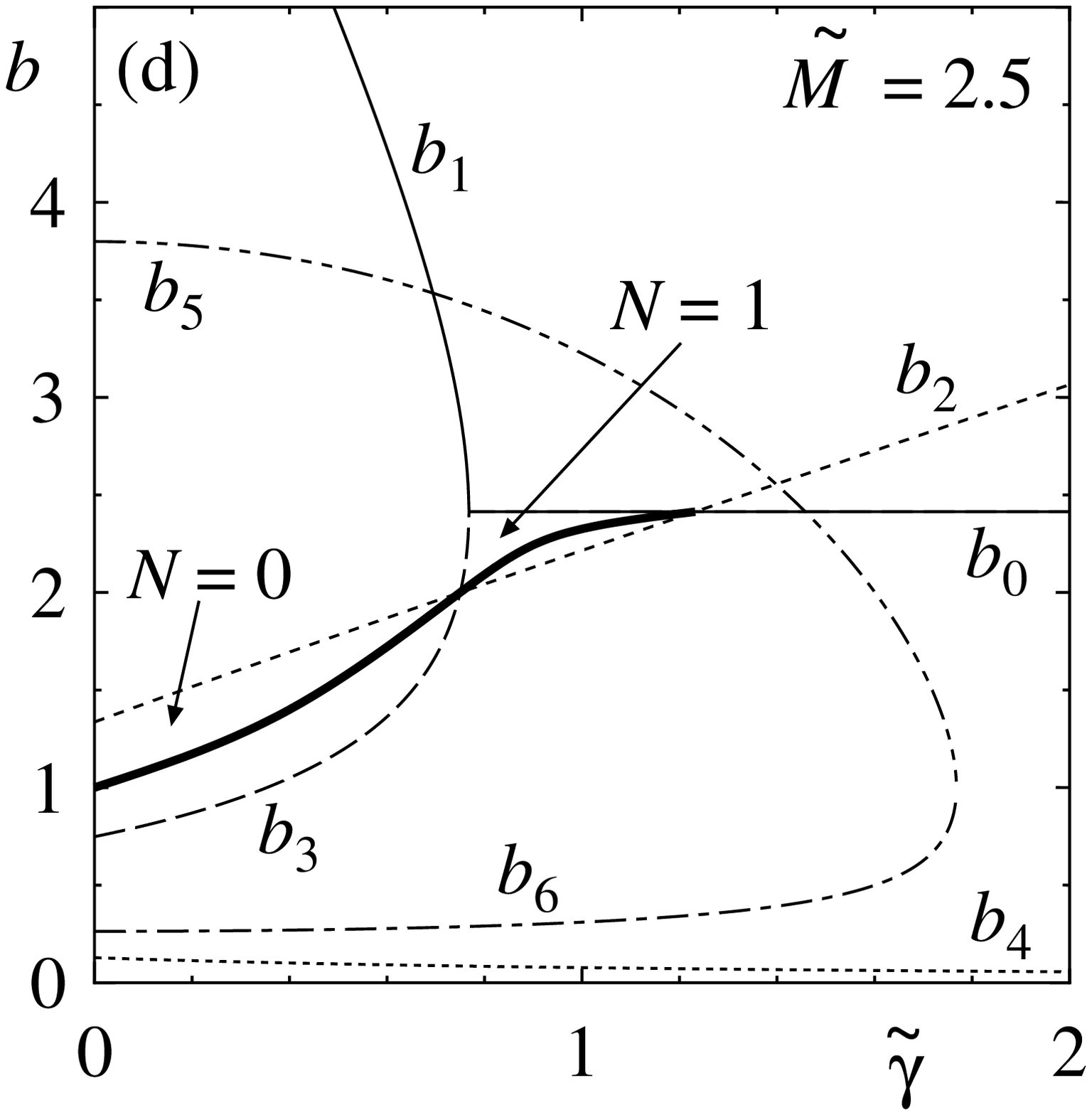}
\end{center}
\end{minipage}
\end{tabular}
\begin{tabular}{cc}
\begin{minipage}{0.5\hsize}
\begin{center}
\hspace{-10mm}
\includegraphics[height=4.0cm]{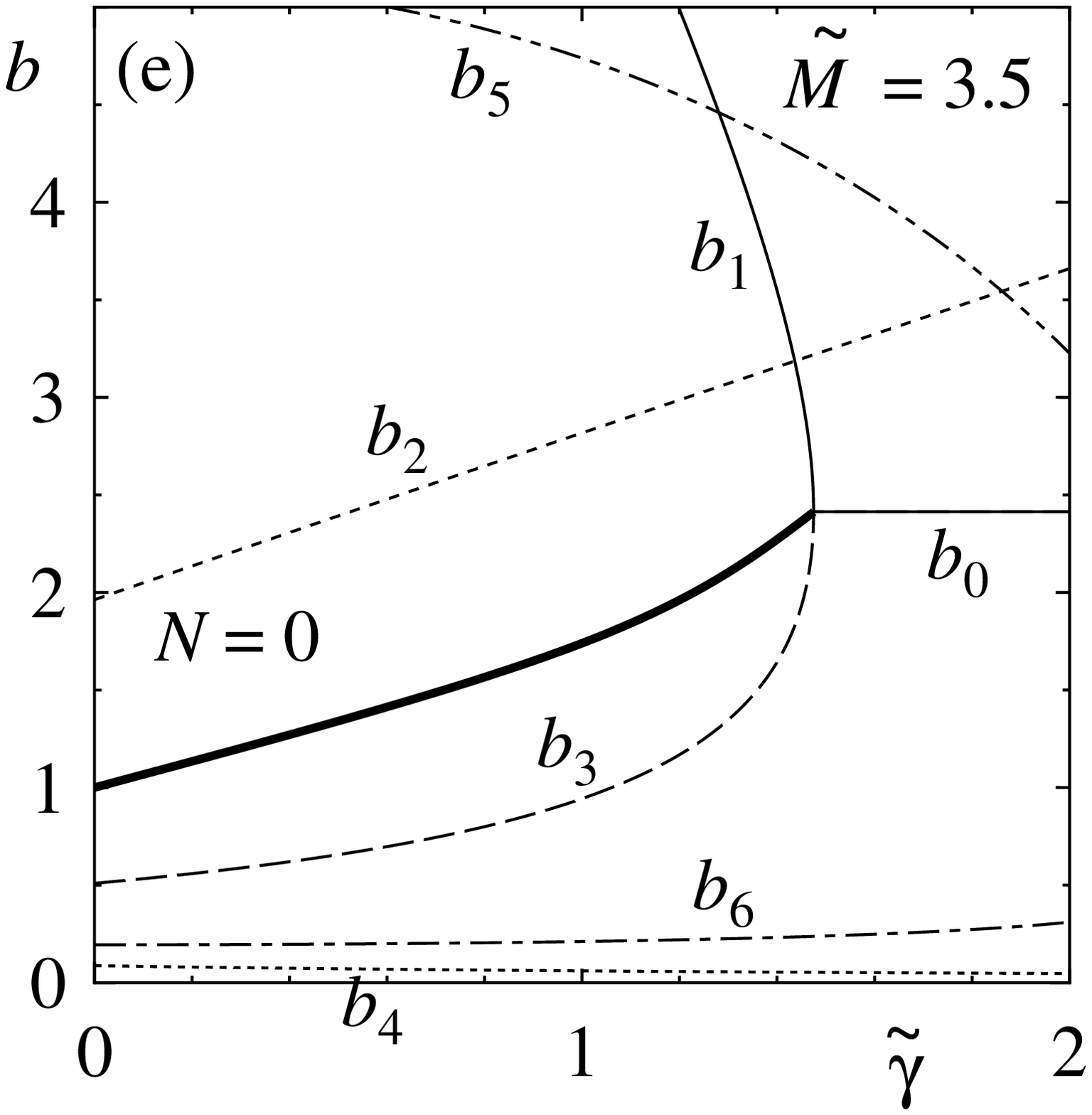}
\end{center}
\end{minipage}
\begin{minipage}{0.5\hsize}
\begin{center}
\hspace{-5mm}
\end{center}
\end{minipage}
\end{tabular}
\caption{
Phase diagrams in the bulk geometry in the $\tilde{\gamma} b$-plane for
$\tilde{M} =$ (a) $0.5$, (b) $1.2$, (c) $2.2$, (d) $2.5$, and (e) $3.5$.
In each panel, the thick solid line represents the trajectory of
$b(\tilde{\gamma})$.
}
\end{figure}

By using the phase diagrams shown in Fig. 1,
let us consider which of the three phases appears in the boundary geometry
under the obc: $\psi_{\sigma}(0,n) = \psi_{\sigma}(L+1,n)
= \psi_{\sigma}(m,0) = \psi_{\sigma}(m,L+1) = 0$.
In the Hermitian limit of $\tilde{\gamma} = 0$, a phase realized
in the boundary geometry is governed by $N$ at $b = 1$,
which corresponds to the pbc.
If $N = 1$ ($N = 0$) at $b = 1$, the nontrivial (trivial) phase
is realized in the boundary geometry.
To extend this bulk--boundary correspondence to the non-Hermitian regime
of $\tilde{\gamma} > 0$, we need to determine $b$ as a function of
$\tilde{\gamma}$ such that $N(\tilde{\gamma},b)$ is
in one-to-one correspondence with a phase realized in the boundary geometry.
A recipe in Ref.~\citen{imura2} tells us that $b(\tilde{\gamma})$
should obey the following three requirements.
Firstly, $b(0) = 1$ at the Hermitian limit of $\tilde{\gamma} = 0$.
Secondly, $b(\tilde{\gamma})$ is allowed to cross gap closing lines $b_{i}$
($i = 0,1,\dots,6$) only at a crossing point between the two.
Thirdly, except at a crossing point, $b(\tilde{\gamma})$ is arbitrary
in a region between two neighboring gap closing lines
in which the two bands are gapped
and therefore $N(\tilde{\gamma},b)$ does not depend on $b = b(\tilde{\gamma})$.

The second requirement
is intuitively understood as follows.
If $b(\tilde{\gamma})$ is on $b_{i}$, a zero-energy solution appears,
giving rise to a gapless spectrum in the bulk geometry.
Consequently, to verify the bulk--boundary correspondence,
the corresponding spectrum in the boundary geometry must also be gapless.
A single solution would be insufficient to construct a general solution
compatible with the obc.~\cite{yao1,yokomizo1}
A crossing point between two $b_{i}$ yields two zero-energy solutions,
constituting a general solution that has degrees of freedom
greater than that in the case of a single solution.
Such a general solution should be compatible with the obc,
resulting in a gapless spectrum in the boundary geometry.
We expect that $b(\tilde{\gamma})$ is allowed to cross $b_{i}$
only at such a crossing point
without breaking the bulk--boundary correspondence.

Possible trajectories of $b(\tilde{\gamma})$ satisfying the three requirements
are shown in Figs.~1(a)--1(e), where each one starts
from $b(0) = 1$ and ends at an entrance to the gapless region.
Except at the initial point and crossing points, $b(\tilde{\gamma})$
is not uniquely determined in accordance with the third requirement.
This does not cause difficulty in the bulk--boundary correspondence
because $N$ is uniquely determined as a function of $\tilde{\gamma}$.
For a given $\tilde{\gamma}$, $N$ at $b(\tilde{\gamma})$ governs
a phase realized in the boundary geometry.
For example, Fig.~1(c) indicates that the phase realized
in the boundary geometry at $\tilde{M} = 2.2$ starts from the trivial phase
($N = 0$) at $\tilde{\gamma} = 0$,
changes to the nontrivial phase ($N = 1$) with increasing $\tilde{\gamma}$,
and finally enters the gapless phase.
From Figs.~1(a)--1(e), we can determine three phase boundaries.
The first one is between the nontrivial region ($N = 1$)
and the gapless region,
the second one is between the trivial region ($N = 0$) and the gapless region,
and the third one is between the trivial region ($N = 0$)
and the nontrivial region ($N = 1$).
Our approach enables us to determine the three boundaries
in a unified manner,
whereas a previous approach~\cite{yao2} was applicable to
only the third boundary.

\begin{figure}[btp]
\begin{center}
\includegraphics[height=4.2cm]{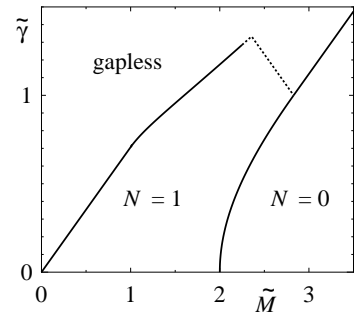}
\end{center}
\caption{
Phase diagram in the boundary geometry, where the regions
designated by $N = 0$ and $N = 1$ respectively correspond to
the topologically trivial and nontrivial phases,
and outside of them is the gapless phase.
The dotted part of the boundary is not confirmed by
spectra in the boundary geometry (see text).
}
\end{figure}
The phase diagram in the boundary geometry (see Fig.~2)
is determined as follows.
Let us focus on the phase boundary between the trivial phase
and the nontrivial phase,
which is at $\tilde{M} = 2$ in the Hermitian limit of $\tilde{\gamma} = 0$.
In the non-Hermitian regime of $\tilde{\gamma} > 0$,
this is determined by the condition of $b_{2}=b_{3}$,
as can be seen from Figs.~1(c) and 1(d).
By solving $b_{2}=b_{3}$ near $\tilde{M} = 2$ in a perturbative manner
with respect to $\tilde{\gamma}$, we find
\begin{align}
  \tilde{M} = 2 + \tilde{\gamma}^{2} -\frac{1}{4}\tilde{\gamma}^{4}
                + \frac{1}{8}\tilde{\gamma}^{6} + \cdots .
\end{align}
This is more accurate than the result reported in Ref.~\citen{yao2}.
We next consider the phase boundary between the trivial and gapless phases.
As can be seen from Fig.~1(e), this is determined by the condition
of $b_{1}=b_{3}$,
resulting in $\tilde{M} = \sqrt{2}\left( \tilde{\gamma}+1 \right)$
when $2\sqrt{2} < \tilde{M}$.
We finally consider the phase boundary
between the nontrivial and gapless phases.
In the range of $\tilde{M} < 1$, this is determined by the condition
of $b_{5}=b_{6}$, as can be seen from Fig.~1(a).
The resulting boundary is given by $\tilde{M} = \sqrt{2}\tilde{\gamma}$.
The phase boundary in the range of $1 < \tilde{M} < \frac{5}{3}\sqrt{2}$
is determined by the condition of $b_{2}=b_{5}$,
as can be seen from Figs.~1(b) and 1(c), while
that in the range of $\frac{5}{3}\sqrt{2} < \tilde{M} < 2\sqrt{2}$ is
determined by the condition of $b_{2}=b_{0}$,
as can be seen from Fig.~1(d).
The resulting boundaries cannot be expressed in a simple form.

The phase boundaries shown in Fig.~2 can be examined by computing
spectra in the boundary geometry under the obc.
The spectra in the case of $\tilde{M} = 0.5$ are shown in Fig.~3,
where the boundary between the nontrivial and gapless phases
is at $\tilde{\gamma}_{\rm c} \approx 0.354$.
The spectrum at $\tilde{\gamma} = 0.34$ is shown in panel (a),
where two series of dots near ${\rm Im}\{E\}=\pm 0.2$ represent
chiral edge states connecting two bulk bands.
The spectrum at $\tilde{\gamma} = 0.36$ is shown in panel (b).
The two bulk bands are separated by a small gap in panel (a), whereas
they are not separated in panel (b), implying that
the system changes from the nontrivial phase to the gapless phase
at $\tilde{\gamma} = \tilde{\gamma}_{\rm c}$.
Spectra in the case of $\tilde{M} = 2.5$ are shown
in Figs.~4(a) and 4(b),
where the boundary between the trivial and nontrivial phases
is at $\tilde{\gamma}_{\rm c} = 0.750$.
The spectrum at $\tilde{\gamma} = 0.74$ is shown in panel (a),
where two bulk bands are separated by a small gap.
The spectrum at $\tilde{\gamma} = 0.76$ is shown in panel (b),
where several dots representing chiral edge states appear in a small gap.
These results imply that the system changes from the trivial phase
to the nontrivial phase at $\tilde{\gamma} = \tilde{\gamma}_{\rm c}$.
Spectra in the case of $\tilde{M} = 3.5$ are shown
in Figs.~4(c) and 4(d),
where the boundary between the trivial and gapless phases
is at $\tilde{\gamma}_{\rm c} \approx 1.475$.
The spectrum at $\tilde{\gamma} = 1.47$ is shown in panel (c),
where two bulk bands are separated by a small gap.
The spectrum at $\tilde{\gamma} = 1.48$ is shown in panel (d),
where a gap closes
and several states appear on the line of ${\rm Re}\{E\} = 0$.
These results imply that the system changes from the trivial phase
to the gapless phase at $\tilde{\gamma} = \tilde{\gamma}_{\rm c}$.
\begin{figure}[btp]
\begin{tabular}{cc}
\begin{minipage}{0.5\hsize}
\begin{center}
\hspace{-10mm}
\includegraphics[height=4.0cm]{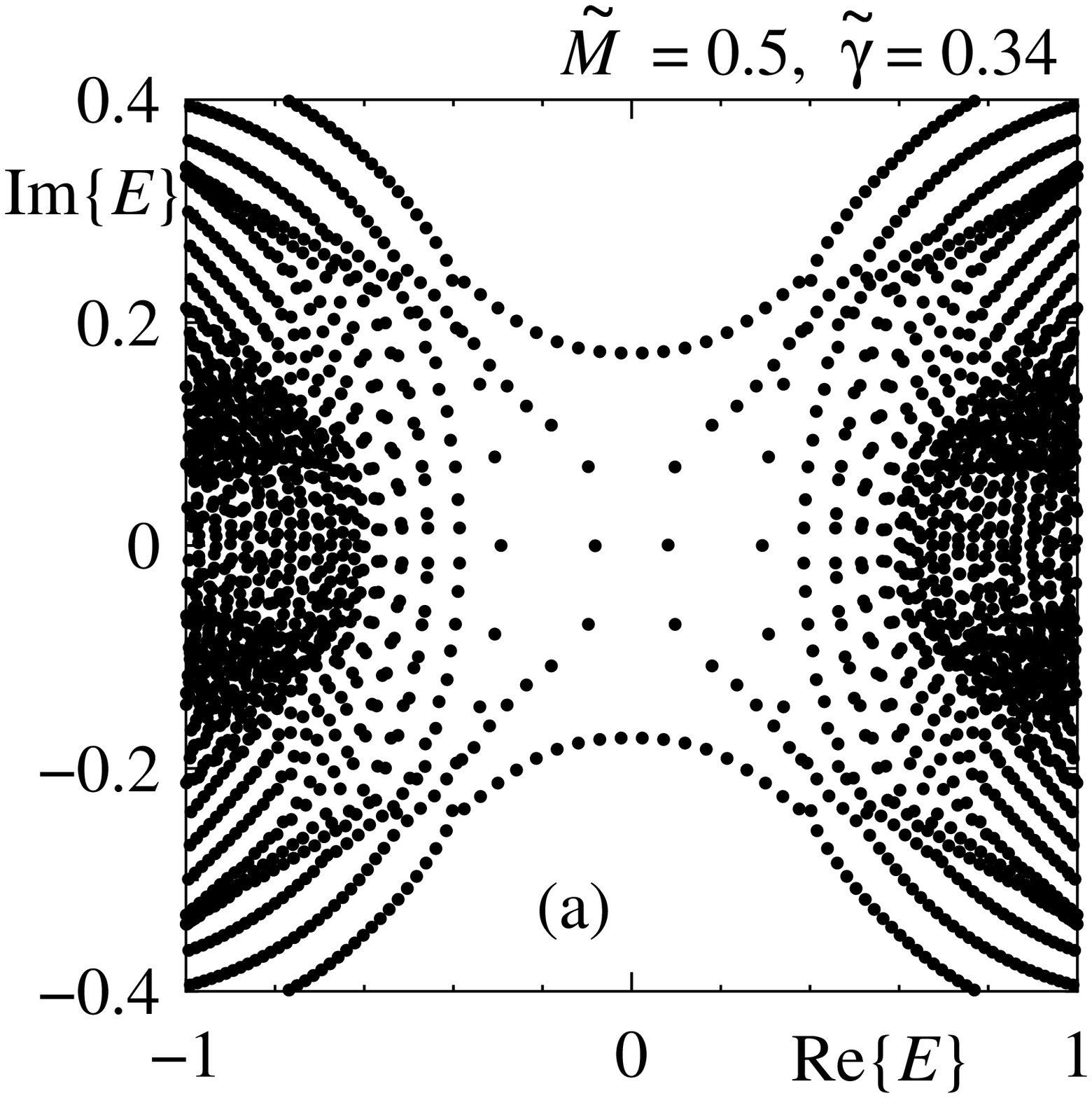}
\end{center}
\end{minipage}
\begin{minipage}{0.5\hsize}
\begin{center}
\hspace{-5mm}
\includegraphics[height=4.0cm]{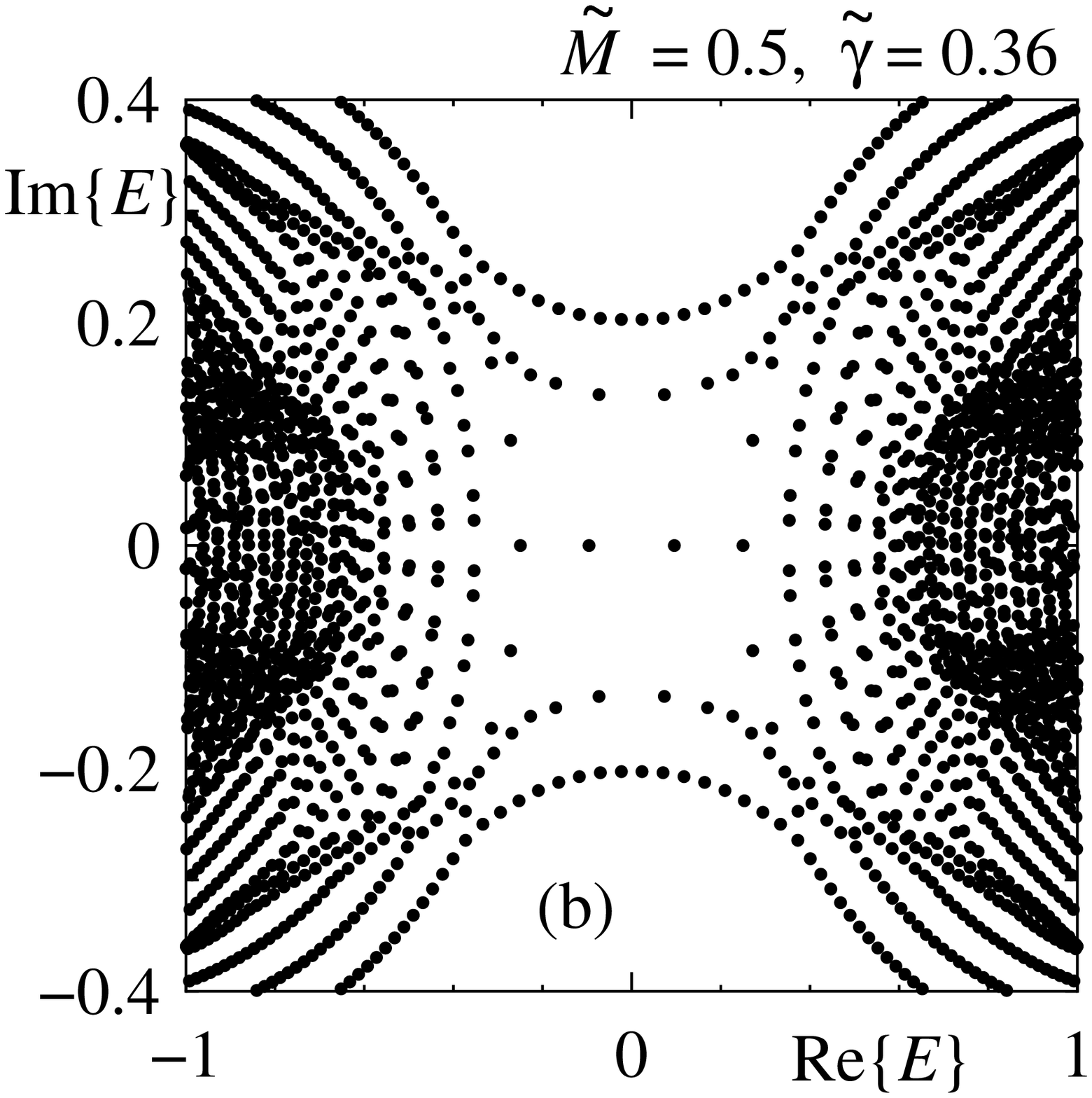}
\end{center}
\end{minipage}
\end{tabular}
\caption{
Spectra in the boundary geometry of $80 \times 80$ sites
at $\tilde{M} = 0.5$ with $\tilde{\gamma} =$ (a) $0.34$ and (b) $0.36$,
where the phase boundary is at $\tilde{\gamma}_{\rm c} \approx 0.354$.
}
\end{figure}
\begin{figure}[btp]
\begin{tabular}{cc}
\begin{minipage}{0.5\hsize}
\begin{center}
\hspace{-10mm}
\includegraphics[height=2.6cm]{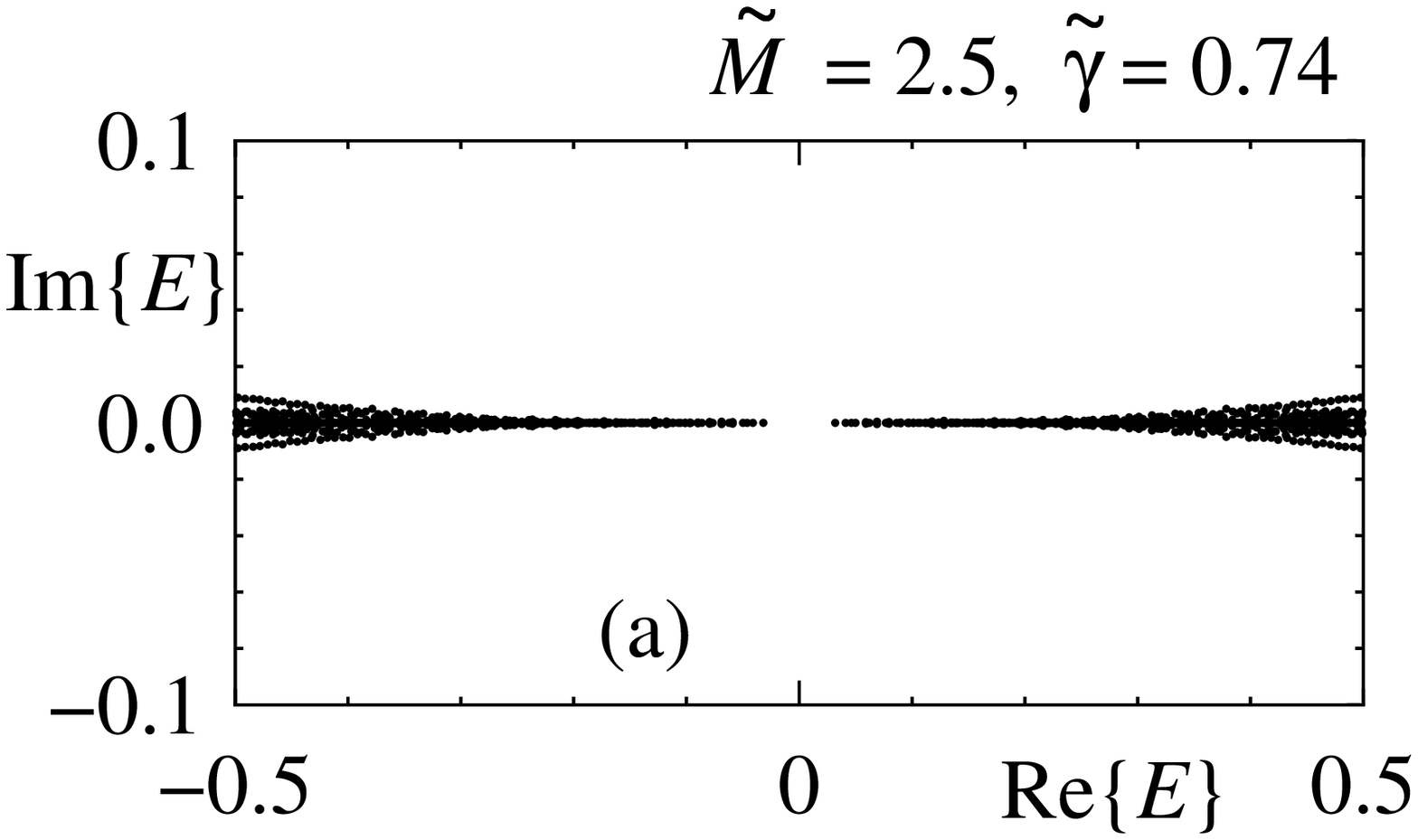}
\end{center}
\end{minipage}
\begin{minipage}{0.5\hsize}
\begin{center}
\hspace{-5mm}
\includegraphics[height=2.6cm]{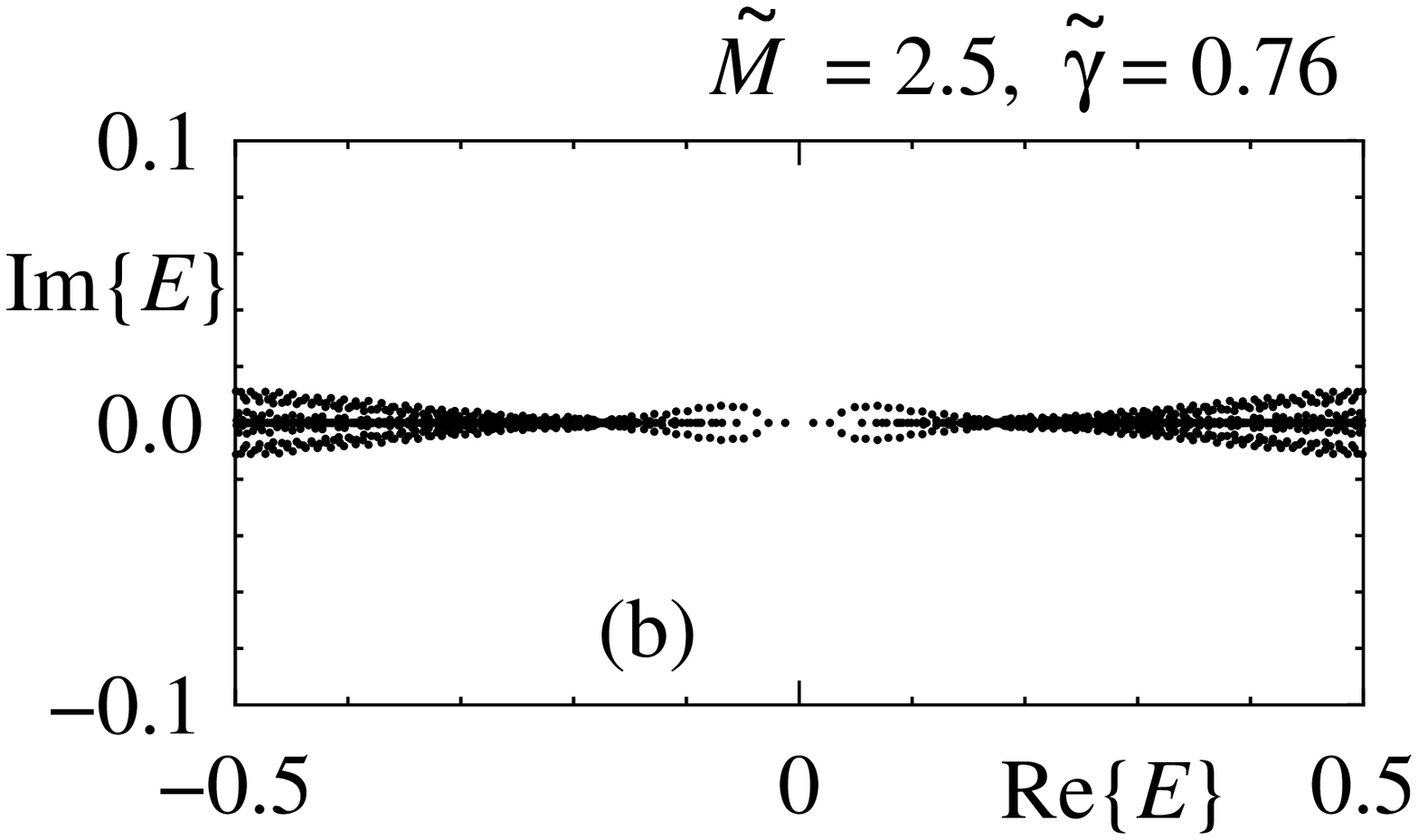}
\end{center}
\end{minipage}
\end{tabular}
\begin{tabular}{cc}
\begin{minipage}{0.5\hsize}
\begin{center}
\hspace{-10mm}
\includegraphics[height=2.6cm]{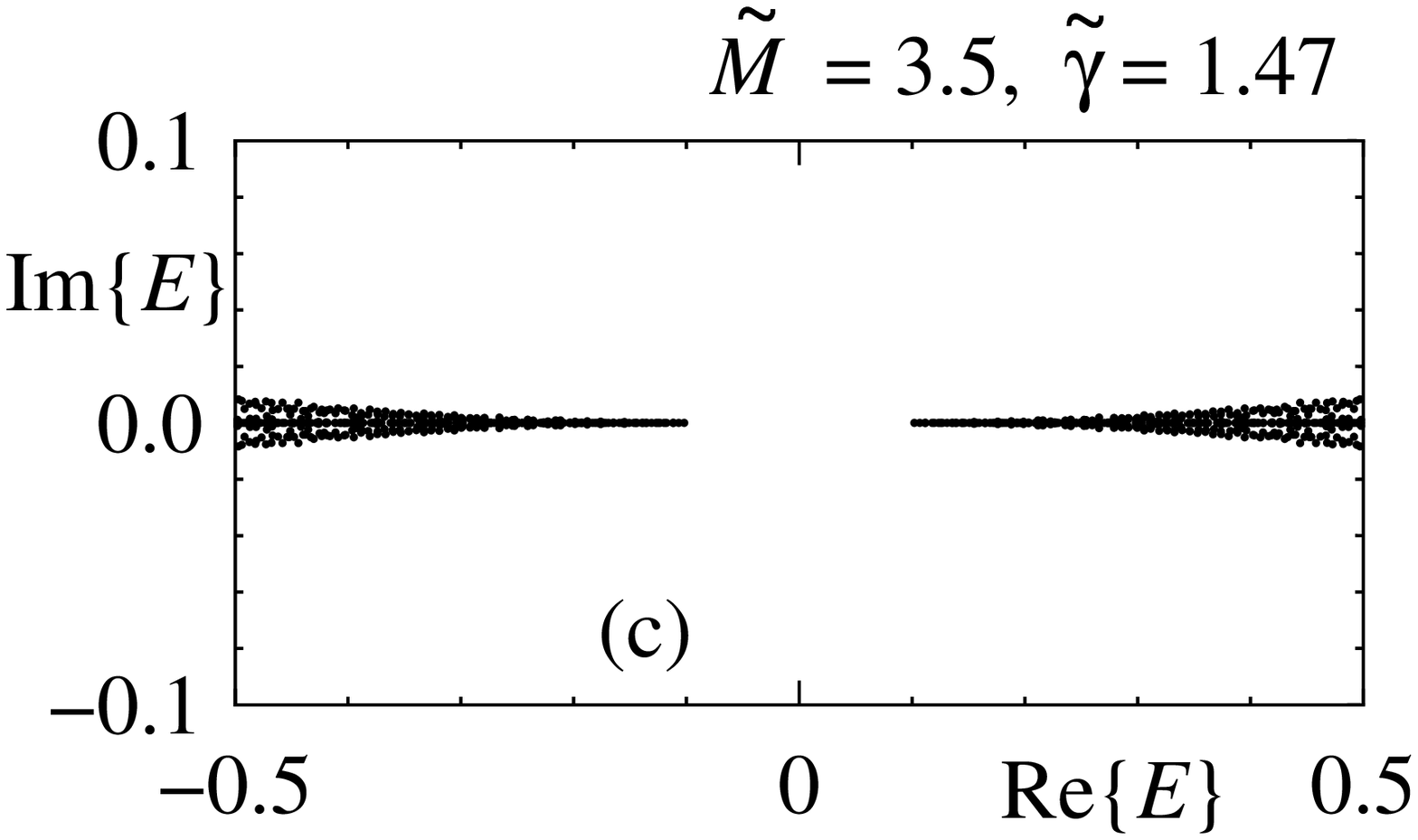}
\end{center}
\end{minipage}
\begin{minipage}{0.5\hsize}
\begin{center}
\hspace{-5mm}
\includegraphics[height=2.6cm]{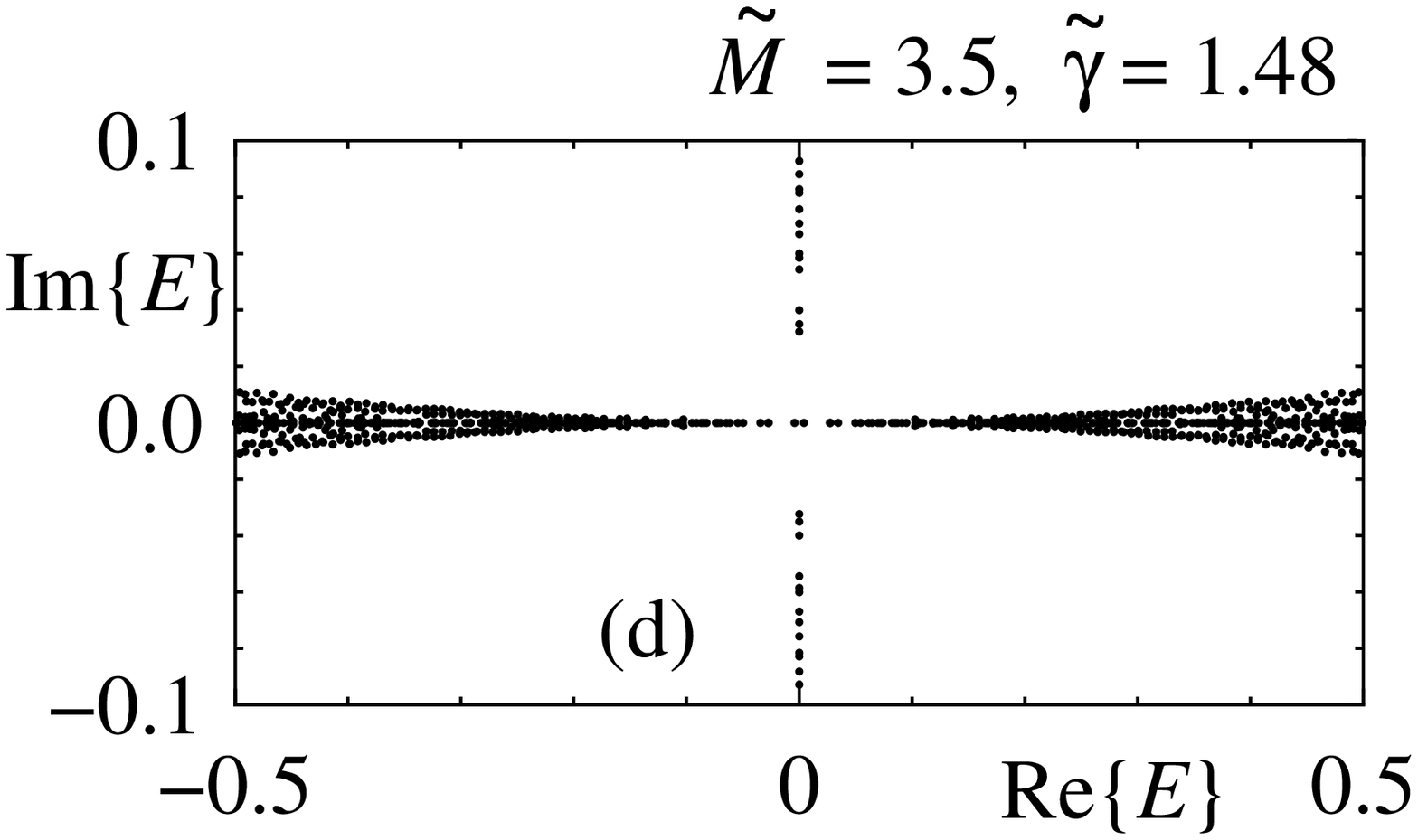}
\end{center}
\end{minipage}
\end{tabular}
\caption{
Spectra in the boundary geometry of $160 \times 160$ sites
at $\tilde{M} = 2.5$
with $\tilde{\gamma} =$ (a) $0.74$ and (b) $0.76$,
where the phase boundary is at $\tilde{\gamma}_{\rm c} = 0.750$,
and those at $\tilde{M} = 3.5$
with $\tilde{\gamma} =$ (c) $1.47$ and (d) $1.48$,
where the phase boundary is at $\tilde{\gamma}_{\rm c} \approx 1.475$.
}
\end{figure}

A part of the boundary represented by the dotted line in Fig.~2
is not confirmed by corresponding spectra in the boundary geometry.
Indeed, a gap between two bulk bands does not close near
this part of the boundary
in the cases of $L = 80$ and $160$ (data not shown).
The reason for this is unclear and is left as an open problem.

In summary, we demonstrated that a scenario of non-Hermitian
bulk--boundary correspondence~\cite{imura1,imura2} is properly applied to
the model of a non-Hermitian Chern insulator in Ref.~\citen{yao2}.
In this model, the phase boundary between trivial and nontrivial phases
has been approximately determined in a rather ad hoc manner.~\cite{yao2}
By applying the scenario, we accurately determined this phase boundary
as well as two other boundaries that were not revealed previously.
Two problems remain to be addressed:
rigorous verification of the second requirement for $b(\tilde{\gamma})$
and confirmation of a missing part of the phase boundary.

\section*{Acknowledgment}

This work was supported by JSPS KAKENHI Grant Number JP18K03460.


\begin{thebibliography}{99}

\bibitem{thouless} D. J. Thouless, M. Kohmoto, P. Nightingale,
and M. den Nijs, Phys. Rev. Lett. {\bf 49}, 405 (1982).

\bibitem{kohmoto} M. Kohmoto, Ann. Phys. {\bf 160}, 343 (1985).

\bibitem{kane} C. L. Kane and E. J. Mele,
Phys. Rev. Lett. {\bf 95}, 146802 (2005).

\bibitem{fu1} L. Fu, C. L. Kane, and E. J. Mele,
Phys. Rev. Lett. {\bf 98}, 106803 (2007).

\bibitem{moore} J. E. Moore and L. Balents,
Phys. Rev. B {\bf 75}, 121306 (2007).

\bibitem{roy} R. Roy, Phys. Rev. B {\bf 79}, 195322 (2009).

\bibitem{ryu1} S. Ryu, A. P. Schnyder, A. Furusaki, and A. W. W. Ludwig,
New J. Phys. {\bf 12}, 065010 (2010).

\bibitem{fu2} L. Fu, Phys. Rev. Lett. {\bf 106}, 106802 (2011).

\bibitem{hasan} M. Z. Hasan and C. L. Kane,
Rev. Mod. Phys. {\bf 82}, 3045 (2010).

\bibitem{qi} X.-L. Qi and S.-C. Zhang,
Rev. Mod. Phys. {\bf 83}, 1057 (2011).

\bibitem{hatsugai} Y. Hatsugai, Phys. Rev. Lett. {\bf 71}, 3697 (1993).

\bibitem{ryu2} S. Ryu and Y. Hatsugai,
Phys. Rev. Lett. {\bf 89}, 077002 (2002).

\bibitem{hatano} N. Hatano and D. R. Nelson,
Phys. Rev. Lett. {\bf 77}, 570 (1996).

\bibitem{bender} C. M. Bender and S. Boettcher,
Phys. Rev. Lett. {\bf 80}, 5243 (1998).

\bibitem{brody} D. C. Brody, J. Phys. A {\bf 47}, 035305 (2014).

\bibitem{esaki} K. Esaki, M. Sato, K. Hasebe, and M. Kohmoto,
Phys. Rev. B {\bf 84}, 205128 (2011).

\bibitem{t_lee} T. E. Lee, Phys. Rev. Lett. {\bf 116}, 133903 (2016).

\bibitem{xu} Y. Xu, S.-T. Wang, and L.-M. Duan,
Phys. Rev. Lett. {\bf 118}, 045701 (2017).

\bibitem{ashida} Y. Ashida, S. Furukawa, and M. Ueda,
Nat. Commun. {\bf 8}, 15791 (2017).

\bibitem{xiong} Y. Xiong, J. Phys. Commun. {\bf 2}, 035043 (2018).

\bibitem{gong} Z. Gong, Y. Ashida, K. Kawabata, K. Takasan, S. Higashikawa,
and M. Ueda, Phys. Rev. X {\bf 8}, 031079 (2018).

\bibitem{kunst1} F. K. Kunst, E. Edvardsson, J. C. Budich, and E. J. Bergholtz,
Phys. Rev. Lett. {\bf 121}, 026808 (2018).

\bibitem{yao1} S. Yao and Z. Wang, Phys. Rev. Lett. {\bf 121}, 086803 (2018).

\bibitem{yao2} S. Yao, F. Song, and Z. Wang,
Phys. Rev. Lett. {\bf 121}, 136802 (2018).

\bibitem{zyuzin} A. A. Zyuzin and A. Yu. Zyuzin,
Phys. Rev. B {\bf 97}, 041203 (2018).

\bibitem{alvarez} V. M. Martinez Alvarez, J. E. Barrios Vargas,
and L. E. F. Foa Torres, Phys. Rev. B {\bf 97}, 121401 (2018).

\bibitem{yoshida1} T. Yoshida, R. Peters, and N. Kawakami,
Phys. Rev. B {\bf 98}, 035141 (2018).

\bibitem{kawabata1} K. Kawabata, K. Shiozaki, and M. Ueda
Phys. Rev. B {\bf 98}, 165148 (2018).

\bibitem{kawabata2} K. Kawabata, S. Higashikawa, Z. Gong, Y. Ashida,
and M. Ueda, Nat. Commun. {\bf 10}, 297 (2019).

\bibitem{longhi1} S. Longhi, Phys. Rev. Res. {\bf 1}, 023013 (2019).

\bibitem{yokomizo1} K. Yokomizo and S. Murakami,
Phys. Rev. Lett. {\bf 123}, 066404 (2019).

\bibitem{okuma1} N. Okuma and M. Sato,
Phys. Rev. Lett. {\bf 123}, 097701 (2019).

\bibitem{song} F. Song, S. Yao, and Z. Wang,
Phys. Rev. Lett. {\bf 123}, 246801 (2019).

\bibitem{herviou} L. Herviou, J. H. Bardarson, and N. Regnault,
Phys. Rev. A {\bf 99}, 052118 (2019).

\bibitem{okugawa} R. Okugawa and T. Yokoyama,
Phys. Rev. B {\bf 99}, 041202 (2019).

\bibitem{yoshida2} T. Yoshida, R. Peters, N. Kawakami, and Y. Hatsugai,
Phys. Rev. B {\bf 99}, 121101 (2019).

\bibitem{c_lee} C. H. Lee and R. Thomale, Phys. Rev. B {\bf 99}, 201103 (2019).

\bibitem{papaj} M. Papaj, H. Isobe, and L. Fu,
Phys. Rev. B {\bf 99}, 201107 (2019).

\bibitem{kunst2} F. K. Kunst and V. Dwivedi,
Phys. Rev. B {\bf 99}, 245116(2019).

\bibitem{imura1} K.-I. Imura and Y. Takane,
Phys. Rev. B {\bf 100}, 165430 (2019).

\bibitem{zhou1} L. Zhou, Phys. Rev. B {\bf 100}, 184314 (2019).

\bibitem{torres} L. E. F. Foa Torres,
J. Phys. Mater. {\bf 3}, 014002 (2019).

\bibitem{borgnia} D. S.  Borgnia, A. J. Kruchkov, and R.-J. Slager,
Phys. Rev. Lett. {\bf 124}, 056802 (2020).

\bibitem{okuma2} N. Okuma, K. Kawabata, K. Shiozaki, and M. Sato,
Phys. Rev. Lett. {\bf 124}, 086801 (2020).

\bibitem{zhang} K. Zhang, Z. Yang, and C. Fang,
Phys. Rev. Lett. {\bf 125}, 126402 (2020).

\bibitem{yi} Y. Yi and Z. Yang,
Phys. Rev. Lett. {\bf 125}, 186802 (2020).

\bibitem{yang} Z. Yang, K. Zhang, C. Fang, and J. Hu,
Phys. Rev. Lett. {\bf 125}, 226402 (2020),

\bibitem{yokomizo2} K. Yokomizo and S. Murakami,
Phys. Rev. Res. {\bf 2}, 043045 (2020).

\bibitem{zhou2} L. Zhou, Phys. Rev. B {\bf 101}, 014306 (2020).

\bibitem{e_lee} E. Lee, H. Lee, and B.-J. Yang,
Phys. Rev. B {\bf 101}, 121109 (2020).

\bibitem{wu1} H. Wu and J.-H. An,
Phys. Rev. B {\bf 102}, 041119 (2020).

\bibitem{wu2} H. C. Wu, X. M. Yang, L. Jin, and Z. Song,
Phys. Rev. B {\bf 102}, 161101 (2020).

\bibitem{longhi2} S. Longhi,
Phys. Rev. B {\bf 102}, 201103 (2020).

\bibitem{kawabata3} K. Kawabata, M. Sato, and K. Shiozaki,
Phys. Rev. B {\bf 102}, 205118 (2020).

\bibitem{mochizuki} K. Mochizuki, D. Kim, N. Kawakami, and H. Obuse,
Phys. Rev. A {\bf 102}, 062202 (2020).

\bibitem{li} L. Li, C.-H. Lee, S. Mu, and J. Gong,
Nat. Commun. {\bf 11}, 5491 (2020).

\bibitem{koch} R. Koch and J. C. Budich, Eur. Phys. J. D {\bf 74}, 70 (2020).

\bibitem{yuce} C. Yuce, Ann. Phys. (NY) {\bf 415}, 168098 (2020).

\bibitem{mostafavi} F. Mostafavi, C. Yuce, O. S. Magan\~{a}-Loaiza,
H. Schomerus, and H. Ramezani, Phys. Rev. Res. {\bf 2}, 032057 (2020).

\bibitem{wang} X.-R. Wang, C.-X. Guo, Q. Du, and S.-P. Kou,
Chin. Phys. Lett. {\bf 37} 117303 (2020).

\bibitem{zhao} X.-L. Zhao, L.-B. Chen, L.-B. Fu, and X.-X. Yi,
Ann. Phys. {\bf 532}, 1900402 (2020).

\bibitem{he} Y. He and C.-C. Chien,
J. Phys.: Condens. Matter {\bf 33}, 085501 (2021).

\bibitem{yokomizo3} K. Yokomizo and S. Murakami,
Prog. Theor. Exp. Phys. {\bf 2020}, 12A102 (2020).

\bibitem{imura2} K.-I. Imura and Y. Takane,
Prog. Theor. Exp. Phys. {\bf 2020}, 12A103 (2020).

\bibitem{kondo} H. Kondo, Y. Akagi, and H. Katsura,
Prog. Theor. Exp. Phys. {\bf 2020}, 12A104 (2020).

\bibitem{kawasaki} M. Kawasaki, K. Mochizuki, N. Kawakami, and H. Obuse,
Prog. Theor. Exp. Phys. {\bf 2020}, 12A105 (2020).

\bibitem{yoshida3} T. Yoshida, R. Peters, N. Kawakami, and Y. Hatsugai,
Prog. Theor. Exp. Phys. {\bf 2020}, 12A109 (2020).



\end{thebibliography}
\end{document}